\documentclass{blois}

\def\Journal#1#2#3#4{{#1} {\bf #2,} #3 (#4)}

\def\PRL{\em Phys. Rev. Lett.}
\def\PRD{{\em Phys. Rev.} D}

\def\JHEP{{\em JHEP}}
\def\Chin{{\em Chin. Phys.} C}

\def\be{\begin{equation}}
\def\ee{\end{equation}}
\def\bea{\begin{eqnarray}}
\def\eea{\end{eqnarray}}

\newcommand{\Photo}{\includegraphics[height=35mm]{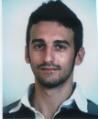}}

\begin{document}
\title{NEXT-TO-NEXT-LEADING-ORDER ANALYSIS OF \\ ELECTROWEAK VACUUM STABILITY \\ AND RISING INFLECTION POINT}

\author{G. IACOBELLIS}

\address{Department of Physics and Earth Science and INFN, \\ via Saragat 1, 44122, Ferrara, Italy}

\maketitle\abstracts{In this talk, based on the work~\cite{iac}, we show an analysis on the gauge-independent observables associated with two stationary configurations of the Standard Model (SM) potential (extrapolated to high energy at Next-to-Next-to-Leading-Order (NNLO)): i) the value of the top mass ensuring stability of the SM electroweak vacuum and ii) the value of the Higgs potential at a rising inflection point. We examine in detail and reappraise the experimental and theoretical uncertainties which plague their determination, keeping alive the possibility for the SM of being stable and studying applications of such configuration to models of primordial inflation.}

\section{Calculation}

The stability issue of the SM and the consequent possible extrapolation of the model up to high energy via the Renormalisation Group Equation (RGE) and without any new physics contribution, represent a very interesting task, being considered the effects induced by the shape of the Higgs potential for very high field values on the dynamics of the early universe. 

Our goal is to carry an updated study (at the present state of the art, namely the NNLO) of the gauge-independent observables associated with two important stationary configurations: the stability bound (two degenerate vacua), ensured by a particular value of the top mass and the case of a rising inflection point, close to the Planck scale.

We adopt the $\overline{MS}$ scheme, consider the matching and the updated three-loops RGE evolution of the relevant couplings (the Higgs quartic coupling $\lambda$, the gauge couplings $g$, $g'$, $g_{3}$, the top Yukawa coupling $h_{t}$ and the anomalous dimension of the Higgs field $\gamma$). We then compute the RGE-improved effective potential. Since in our calculation the boundary conditions are given at the top quark mass value $m_{t}$, we take $\mu=m_{t}$ as the preferred renormalisation scale, in order to minimise the effect of the large logarithms corrections at one loop in the effective potential, improving in such a way the convergence of perturbation theory.

As starting values of the parameters involved in the calculation, we exploit the results of~\cite{bed}, updating them, when necessary, by means of the latest Particle Data Group data~\cite{pdg}. It is worth mentioning that the value used in our work for the strong coupling constant $\alpha_{s}^{(5)}$ at $m_{Z}$ in the five-flavours SM, whose contribution is dominant in the uncertainty of $g_{3}(m_{t})$, has a $1\sigma$ error approximately doubled with respect to the value used in the previous analyses~\cite{deg}$^{,}$ \cite{but}: $\alpha_{s}^{(5, exp)}=0.1181\pm0.0013$. The uncertainty on $\lambda(m_{t})$ is dominated by the experimental error on the Higgs mass ($m_{H}^{exp}=125.09~GeV\pm0.24~GeV$) and the theoretical uncertainty related to the matching procedure (scale variation and truncation). The matching of the Yukawa top quark coupling $h_{t}(m_{t})$ is mainly affected by the experimental error on the top mass along with the usual uncertainty related to the matching procedure itself. The benchmark value is $m_{t}^{exp}=173.34~GeV\pm0.76~GeV$. Another unavoidable source of uncertainty resides in the common identification of the top quark mass $m_{t}$ with the actual measured Monte Carlo (MC) parameter $m_{t}^{MC}$: the translation from the MC mass definition to a theoretically well-defined short distance mass at low energy is estimated to be of the order of $1~GeV$ (according to the most conservative groups), or something comparable to a $100~MeV$ discrepancy.

The last ingredient for a complete and reliable calculation, as already anticipated, is the RGE-improved effective potential: we consider the one-loop radiative correction (Coleman-Weinberg) to the SM Higgs potential plus the two-loop contribution, taken in the compact form proposed by~\cite{but}. One problem in this treatment is related to the fact that when $\lambda(t)$ becomes negative, the Higgs and Goldstone contributions turn out to be small, but complex: some authors~\cite{and} showed that the procedure of setting to zero the mentioned terms~\cite{deg}$^{,}$ \cite{but} is theoretically justified when $\lambda$ is small: a new resummation technique should be used, and the dangerous ghosts have to be accounted to higher-order terms.

%

Nevertheless one has to be aware that the truncation of the effective potential loop expansion at some loop order, introduces a theoretical error. For this sake, it is useful to define the parameter $\alpha$ via $\mu(t)=\alpha\phi(t)$, where $\mu$ is the renormalisation scale and $\phi$ the Higgs field, and study the dependence of our gauge-invariant observables on $\alpha$. The higher the loop order expansion, the less the dependence on $\alpha$.

\section{Criticality}

\begin{figure}
\begin{minipage}{0.50\linewidth}
\centerline{\includegraphics[width=0.9\linewidth]{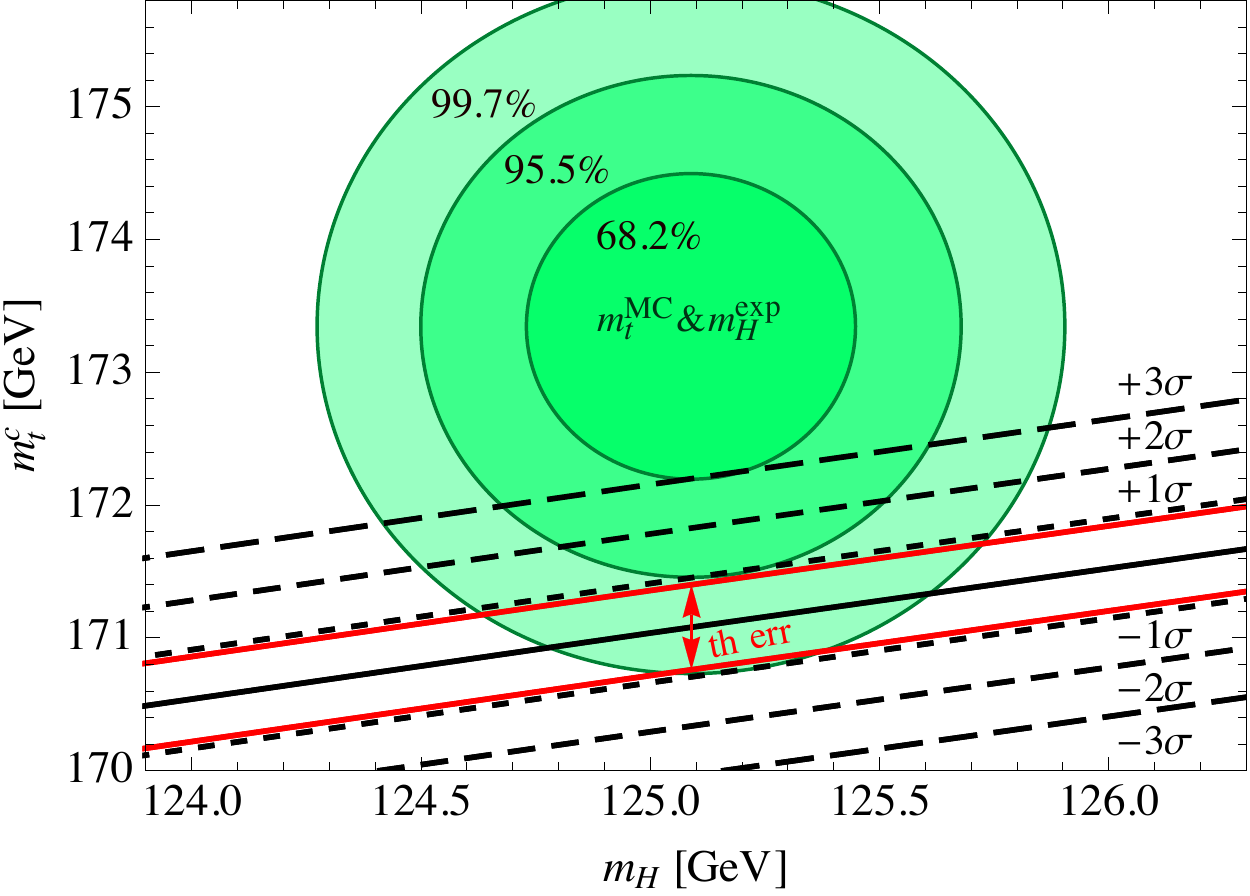}}
\end{minipage}
\begin{minipage}{0.50\linewidth}
\centerline{\includegraphics[width=0.9\linewidth]{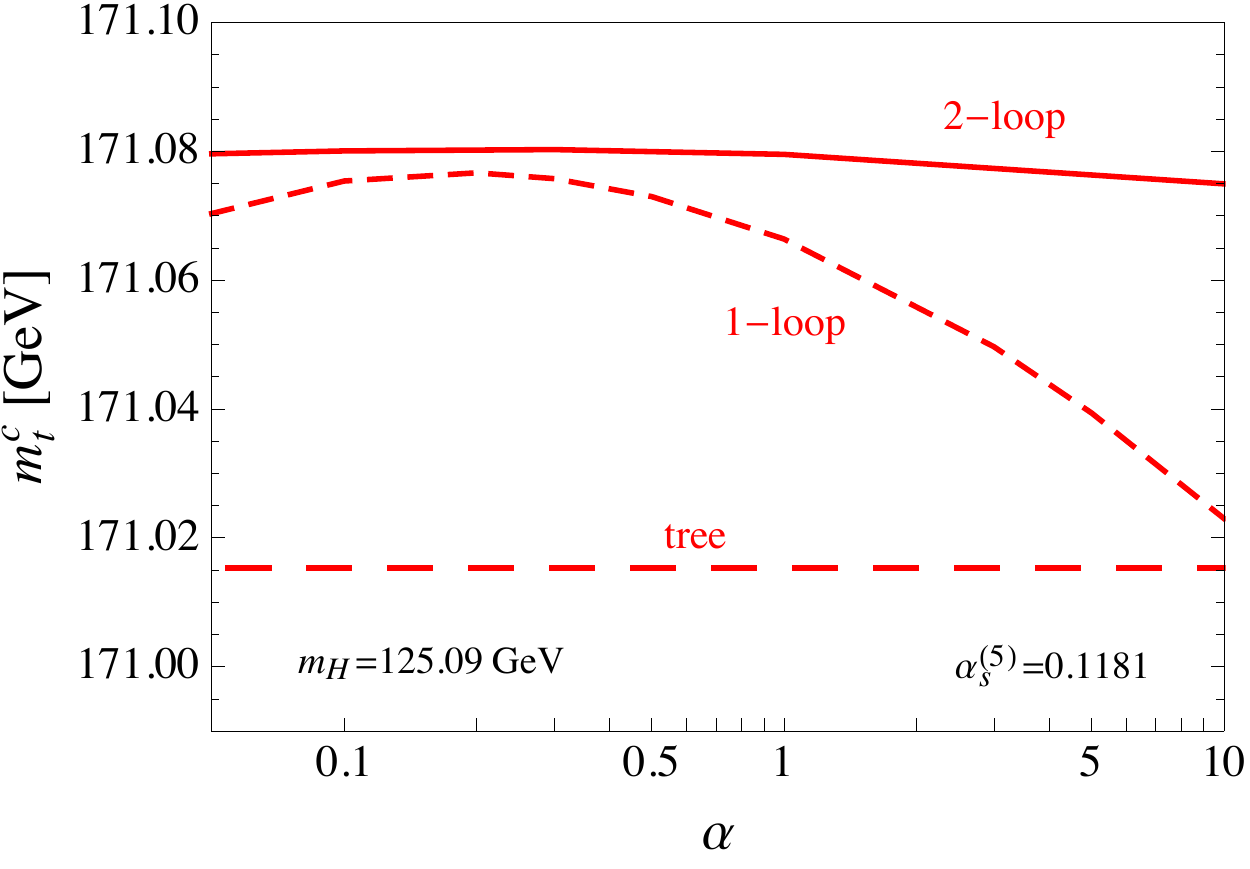}}
\end{minipage}
\caption[]{\emph{Left panel}: Lines for which the Higgs potential develops a second degenerate minimum at high energy. The solid line corresponds to the central value of $\alpha_{s}^{(5)}$; the dashed lines are obtained by varying $\alpha_{s}^{(5)}$ in its experimental range, up to $3\sigma$. The (red) arrow represents the theoretical error in the position of the lines. The (green) shaded regions are the covariance ellipses obtained combining $m_{t}^{MC}=173.34\pm0.76~GeV$ and $m_{H}^{exp}=125.09\pm0.24~GeV$; the probability of finding $m_{t}^{MC}$ and $m_{H}$ inside the inner (central, outer) ellipse is equal to $68.2\%$ ($95.5\%$, $99.7\%$). \emph{Right panel}: Dependence of $m_{t}^{c}$ on $\alpha$, for increasing level of the effective potential expansion. The central values for $m_{H}$ and $\alpha_{s}^{(5)}$ were used.}
\label{fig:one}
\end{figure}

The $m_{t}^{c}$ value is plagued by experimental and theoretical errors, which we analysed in detail.

The experimental error is the one associated to the precision with which we know the input parameters at the matching scale $m_{t}$. The uncertainty on $m_{t}^{c}$ due to varying $\alpha_{s}^{(5)}$ in its $1\sigma$ range is $0.37~GeV$, while the uncertainty due to varying $m_{H}$ in its $1\sigma$ range is $0.12~GeV$. This can be seen in fig.~\ref{fig:one}, where $m_{t}^{c}$ is displayed as a function of $m_{H}$ for particular values of $\alpha_{s}^{(5)}$. In particular, the solid line refers to its central value, while the dotted, short and long dashed lines refer to the $1\sigma$, $2\sigma$ and $3\sigma$ deviations respectively. In the region below (above) the line the potential is stable (metastable). 

Theoretical errors comes from the approximations done in the three steps of the calculation: matching at $m_t$, running from $m_t$ up to high scales, effective potential expansion. Combining in quadrature the theoretical error in the matching procedure of $\lambda$ and $h_{t}$, we obtain a global uncertainty on $m_{t}^{c}$ due to NNLO matching of about $\pm0.32~GeV$, as represented by the shift (the red arrow) in fig.~\ref{fig:one}. The error committed in the choice of the order of the $\beta$-functions in the RGE is equivalent to the order of magnitude of the subsequent correction: something of the order $10^{-5}~GeV$, thus negligible. The last source of uncertainty on $m_{t}^{c}$ is associated to the effective potential order expansion, estimated by the dependence on the parameter $\alpha$ (see fig.~\ref{fig:one}). Keeping all the parameters fixed at their central values, we let vary $\alpha$ in the interval $0.1-10$: we note that, being the renormalisation scale dependent on $\alpha$ through $\mu(t)=\alpha\phi(t)$, at tree level, the Higgs potential does not depend explicitly on $\mu(t)$, but only implicitly via the running couplings. For this reason the dependence at tree level is negligible, while at the one-loop and the two-loop order is explicit, but again small. 

Summarising, the critical top mass value turns out to be:
\begin{equation}
m_{t}^{c}=171.08\pm0.37_{\alpha_{s}}\pm0.12_{m_{H}}\pm0.32_{th}~GeV\,,
\end{equation}
where the first two errors are the $1\sigma$ variations of $\alpha_{s}^{(5)}$ and $m_{H}$: this result updates and improves, being still consistent with, the well-known results in literature~\cite{bed}$^{,}$ \cite{deg}$^{, }$ \cite{but}. As a result, the updated calculation of the experimental and theoretical uncertainties on $m_{t}^{c}$, in addition to the uncertainty in the identification of the MC and pole top masses, lead us to conclude that the configuration with two degenerate vacua is at present compatible with the experimental data. It is clear that, in order to discriminate in a robust way between stability and metastability, it would be crucial to reduce the experimental uncertainties in both $m_{t}$ and $\alpha_{s}^{(5)}$ . A reduction of the theoretical error in the matching would also be welcome.

\section{Inflection point}

\begin{figure}
\begin{minipage}{0.50\linewidth}
\centerline{\includegraphics[width=0.9\linewidth]{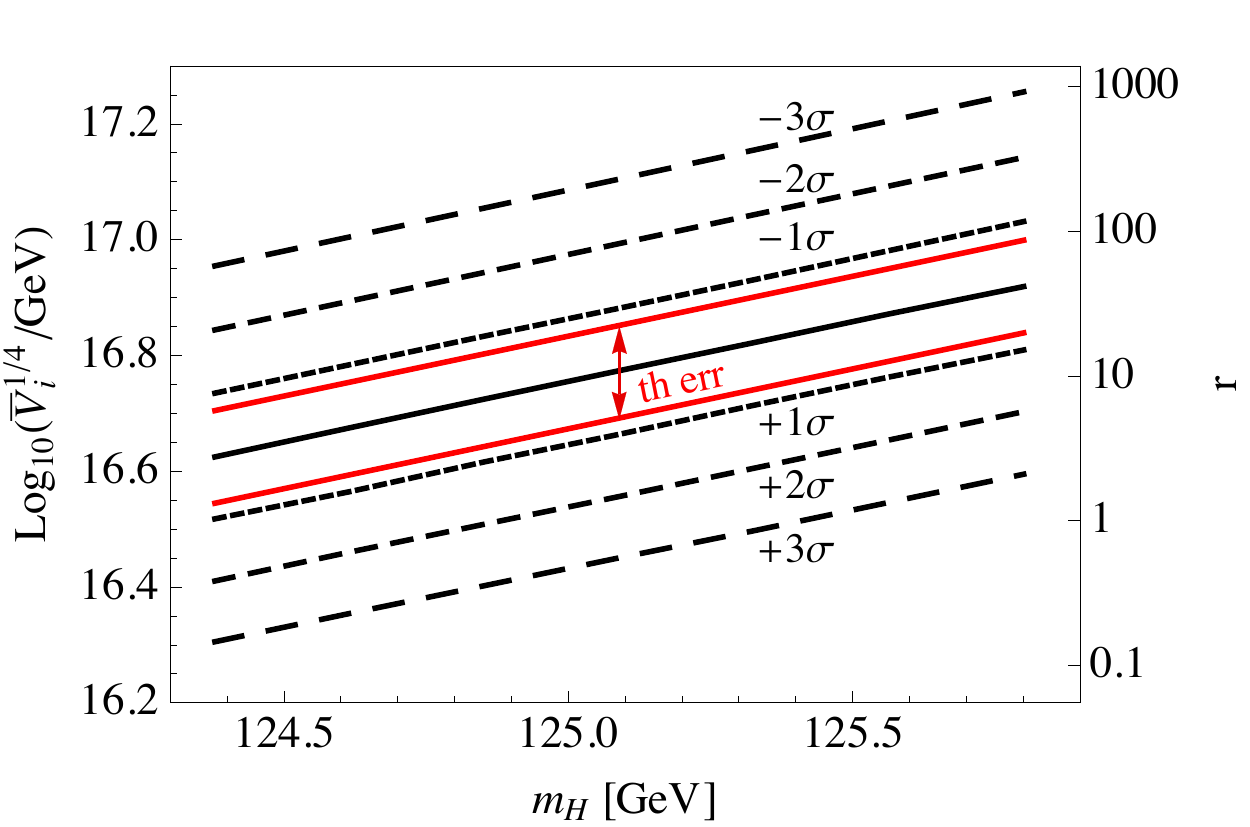}}
\end{minipage}
\begin{minipage}{0.50\linewidth}
\centerline{\includegraphics[width=0.9\linewidth]{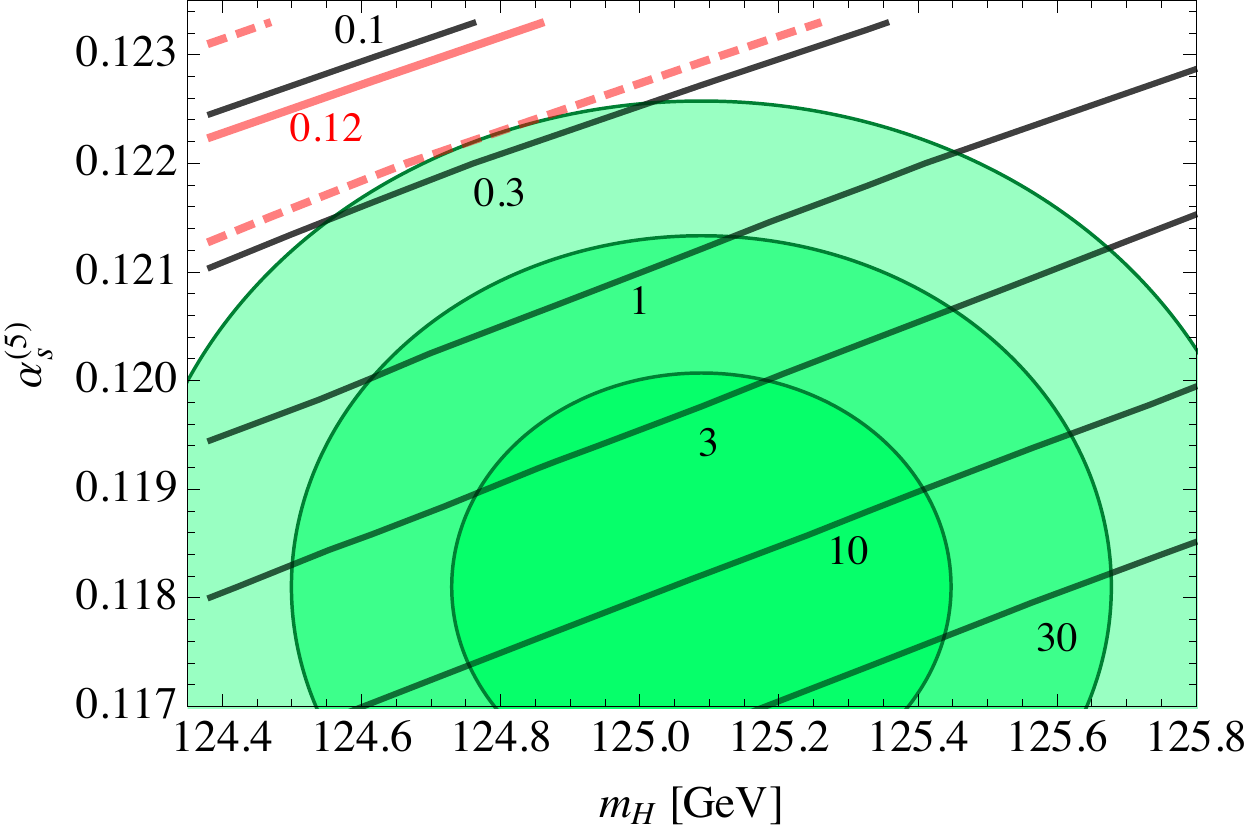}}
\end{minipage}
\caption[]{\emph{Left panel}: Dependence of $\bar{V}_{i}^{1/4}$ on $m_{H}$ for fixed values of $\alpha_{s}^{(5)}$. The (red) arrow and solid lines show the theoretical error due to the matching of $\lambda$. The right vertical axis displays the associated value of the tensor-to-scalar ratio $r$. \emph{Right panel}: Contour levels of $r$ in the plane ($m_{H}$, $\alpha_{s}^{(5)}$). The theoretical uncertainty corresponding to $r = 0.12$ is shown by means of the (red) dashed lines. The shaded regions are the covariance ellipses indicating that the probability of finding the experimental values of $m_{H}$ and $\alpha_{s}^{(5)}$ inside the ellipses are respectively $68.2\%$, $95.4\%$, $99.7\%$.}
\label{fig:two}
\end{figure}

The highness of the effective potential at an inflection point, let us call it $\bar{V}_{i}$, could be directly linked to the ratio
of the scalar to tensor modes of primordial perturbations, $r$, via the relation
\begin{equation}
\bar{V}_{i}=\frac{3\pi^{2}}{2}r A_{s}\,,
\end{equation}
where the amplitude of scalar perturbations is $A_{s}=2.2\times10^{-9}$. It is thus relevant for models of primordial inflation, based on a shallow false minimum~\cite{mas}, to assess the size of the experimental and theoretical errors in the calculation of $\bar{V}_{i}$.

Experimental uncertainties on $\bar V_i$ can be estimated: we let $m_H$ vary in its $3\sigma$ experimental range and, for fixed values of $\alpha_s^{(5)}$, we determine $m^i_t$, the value of the top mass needed to have an inflection point (which is so close to $m^c_t$ that one can read it from the stability line of fig.~\ref{fig:one}). We then evaluate the effective potential at this point, $\bar V_i$. 
The result is displayed in fig.~\ref{fig:two}: one can see that $\bar V_i^{1/4}$ spans one order of magnitude, for decreasing values of $\alpha_s^{(5)}$; the dependence on $m_H$ is less dramatic but still relevant, being about $50\%$.

As done before, the theoretical uncertainties can be divided in three groups: again, the one associated to the matching procedure is the dominant one, with a shift up or down of the lines in fig.~\ref{fig:two} of about $0.08$. The truncation and variation errors are again negligible.

We can conclude that the result of the NNLO calculation is
\begin{equation}
\log_{10} {\bar V}_{i}^{1/4} =  16.77  \pm 0.11_{\alpha_s} \pm 0.05_{m_H}  \pm 0.08_{th}  \,,
\end{equation}
where the first two errors refer the $1\sigma$ variations of $\alpha_s^{(5)}$ and $m_H$ respectively.

In view of an application to inflation, the right axis of fig.~\ref{fig:two} reports the corresponding value of $r$. Notice that the dependence of $r$ on $\alpha_s^{(5)}$ and $m_H$ is extremely strong: when the latter are varied in their $3\sigma$ range, $r$ spans from $0.1$ to $1000$. In addition, the theoretical error in the matching implies an uncertainty on $r$ by a factor of about $2$.

According to the 2015 analysis of the Planck Collaboration, the present upper bound on the tensor to scalar ratio is $r <0.12$ at $95\%$ C.L.~\cite{pla}. This would translate into to the bound $\log_{10} {\bar V}_i^{1/4} < 16.28$ at $95\%$ C.L. and implies a tension at about $3\sigma$ (see fig.~\ref{fig:two}).

The present results are in substantial agreement with those of the most recent detailed analysis on the inflection point configuration, performed by Ballesteros et al.~\cite{bal}.

\section{Conclusions}

Assuming the SM valid up to the Planck scale, we find that stability of the SM is compatible with the present data at the level of $1.5\sigma$. Stability of the SM Higgs potential is thus, in our opinion, still a viable possibility. In order to robustly discriminate between stability and metastability, higher precision measurements of the top quark pole mass, $\alpha_{s}^{(5)}$ and of the matching procedure would be needed. 

As concerns the inflection point configuration, we find that, despite the large theoretical error on the Higgs potential at the inflection point, application of such configuration to models of primordial inflation based on a shallow false minimum displays a $3\sigma$ tension with the recent bounds on the tensor-to-scalar ratio of cosmological perturbations: modifications
due to new physics will be necessarily introduced~\cite{esp}.

\section*{Acknowledgments}

GI wishes to thank the organisers of the 28th Rencontres de Blois for the stimulating atmosphere and for the invitation to give this talk. This work was supported by the PhD grant funded by the Agenzia Spaziale Italiana (ASI). I would acknowledge Isabella Masina as a coauthor of the work~\cite{iac} on which this talk is based.

\end{document}